\documentclass[aps,prb,twocolumn,nofootinbib]{revtex4}  
\usepackage[utf8]{inputenc}
\usepackage[T1]{fontenc}
\usepackage{color}
\usepackage{epsfig,amsfonts}
\usepackage{graphicx}
\usepackage{float} 
\usepackage{comment} 
\usepackage{epstopdf}
\usepackage{blindtext/blindtext}
\usepackage{bbold/bbold}

\def\be{\begin{equation}}
\def\ee{\end{equation}}

\def\bea{\begin{eqnarray}}
\def\eea{\end{eqnarray}}

\def\ben{\begin{enumerate}}
\def\een{\end{enumerate}}

\def\bea{\begin{eqnarray}}
\def\eea{\end{eqnarray}}

\begin{document}

\title{Integrability and dark states of the XX spin-1 central spin model in a transverse field}
\author{Eric De Nadai\footnote{E.D.N and N.M have contributed equally to this work.}, Nathan Maestracci$^*$, Alexandre Faribault}
\affiliation{Universit\'e de Lorraine, CNRS, LPCT, F-54000 Nancy, France}
\begin{abstract}

It was recently shown that, for central spin-1/2 and central spin-1, the XX central spin model is integrable in the presence of a magnetic field oriented perpendicular to the XY plane in which the coupling exists. In the spin-1/2 case, it was also shown, through an appropriate limit of the non-skew symmetric XXZ Richardson-Gaudin models, that it remained integrable even when the magnetic field is tilted to contain an in-plane component.

Although the model has not yet been shown to explicitly belong to a known class of Richardson-Gaudin models, we show, in this work, that the spin-1 case also remains integrable in a titled magnetic field. We do so by writing explicitly the complete set of conserved charges, then showing that these operators obey polynomial relations. It is finally demonstrated numerically that dark states, for which the central spin is completely unentangled with the bath, can emerge at strong enough coupling just as they do in the central spin-1/2 model in an arbitrarily oriented magnetic field.

\end{abstract}

\maketitle

\section{Introduction}

Central spin models describe the interaction of a given specific spin $\vec{S}_0$ with an external magnetic field and a bath of $N$ environmental spins $\vec{S}_j \ \ \forall \ j=1, \dots N$ which do not interact amongst themselves. They have been widely used to describe the coupling of a qubit based on the spin of a single trapped carrier \cite{loss}  with the bath of environmental spins, coupling which ultimately leads to decoherence and to the loss of quantum information \cite{deco1,deco2,deco3}. In the XX limit of the model, which this work is interested in, the couplings to the spins of the bath are restricted to the XY-plane in which they are isotropic so that the hamiltonian reads:
\bea
\hat{H} = \vec{B} \cdot \vec{\hat{S}}_0 +  \sum^{N}_{k = 1}  2 g_k \left(\hat{S}^x_0\hat{S}^x_k + \hat{S}^y_0\hat{S}^y_k\right).
\label{XXwithfield}
\eea

\noindent where the three magnetic field components $B^\alpha$ and the various couplings  $g_k$ are all arbitrary real numbers. For a magnetic field  oriented along the z-axis, $B^x =B^y=0$, the model is rotationally invariant around the z-axis. In this U(1)-symmetric case, it was shown\cite{XXdark} that, when the central spin is chosen to be a spin$-\frac{1}{2}$, the model is integrable and supports dark eigenstates in which the central spin is in the pure state $\left| \uparrow_0 \right>$ or $\left| \downarrow_0 \right>$, i.e. the eigenstates of the $\hat{S}_0^z$ operator. The resulting tensor product eigenstates $\left| \uparrow_0 \right>\otimes\left|\phi_\mathrm{bath}\right>$ (or  $\left| \downarrow_0 \right>\otimes\left|\phi_\mathrm{bath}\right>$) therefore have absolutely no entanglement between the central spin and the bath whose various possible states $\left|\phi_\mathrm{bath}\right>$ can be found through solutions of a set of Bethe equations \cite{XXdark,XXBethe}. Dark states (and dark subspaces of the Hilbert space) can be remarkably desirable since they could provide protected long lived quantum states \cite{control0,control1,control2} in nitrogen-vacancy centers in diamond \cite{laraouinv,dobrovnv,hallnv} or in semiconductor quantum dots \cite{ramsaysemi,hansonsemi,schliemannsemi} for example.

Building on the U(1)-symmetric case\cite{XXdark},  it was shown that dark states can be stable against integrability breaking perturbations \cite{XXperturb}. It was then also demonstrated that, for central spin$-\frac{1}{2}$ the XX-model remains integrable when the applied magnetic field points in an arbitrary direction \cite{dimodark} therefore breaking U(1)-symmetry. The proof is simple since it only relies on taking the appropriate limit $\epsilon_0 \to - j_z$ (with $j_x = j_y$) of the non-skew symmetric elliptic Richardson-Gaudin (R-G) models defined by the set of $N+1$ commuting conserved charges:
\bea
\hat{R}_j =  \vec{B}_j \cdot \vec{\hat{S}}_j + \sum_{\alpha \in \left\{x,y,z\right\}}\left(\sum_{i=0 (\ne j)}^{N} \Gamma^{\alpha}_{i,j} \hat{S}^\alpha_i\hat{S}^\alpha_j +  \Gamma^{\alpha}_{j,j} \hat{S}^\alpha_j\hat{S}^\alpha_j \right).
\label{XYZgeneric}
\eea

\noindent Their mutual commutation and the resulting integrability of the model requires that the various terms be parametrised as \cite{skrypnykquad}:
\bea
B^{\alpha }_j &=& \frac{B_\alpha}{\sqrt{\epsilon_j+j_\alpha}}
\nonumber\\
\Gamma^{\alpha}_{i,j} &=&\frac{\sqrt{\epsilon_j+j_\alpha}\sqrt{\epsilon_i+j_\beta}\sqrt{\epsilon_i+j_\gamma}}{\epsilon_i-\epsilon_j} 
\nonumber\\
\Gamma^{\alpha}_{j,j} &=& \frac{(\epsilon_j+j_\alpha)(2\epsilon_j+j_\beta+j_\gamma)-(\epsilon_j+j_\beta)(\epsilon_j+j_\gamma)}{\sqrt{\epsilon_j+j_\alpha}\sqrt{\epsilon_j+j_\beta}\sqrt{\epsilon_j+j_\gamma}} 
\eea
\noindent where $\gamma\ne\beta\ne\alpha$ form the three distinct direction indices, while $B_\alpha$, $j_\alpha$ and $\epsilon_k$ ($\forall k=0,1 \dots N)$ are $N+7$ arbitrary free parameters chosen real to insure hermiticity. Restricting oneself to the XXZ case ($j_x=j_y$) and taking the limit $j_z \to -\epsilon_0$ (while keeping $B^{z }_0$ finite through a diverging $B_z$ ), the resulting conserved charge $R_0$ does become the XX-hamiltonian (\ref{XXwithfield}). However, it exclusively does so for a spin$-\frac{1}{2}$ realisation of the central spin. Indeed, when $S_0=\frac{1}{2}$, the "self-coupling terms" $\propto S^\alpha_0 S^\alpha_0$ present in $R_0$ reduces to a constant which can simply be ignored by shifting the zero-energy point. However, for any higher spin representation this term remains non-trivial and does not vanish when performing the same limit. For higher spins, the desired XX model (\ref{XXwithfield}) can simply not be obtained as a limit of an elliptic model  (\ref{XYZgeneric}), due to the presence of this additional "self-coupling" term.

Nonetheless, by constructing  the commuting conserved charges and, via a Bethe Ansatz, the explicit eigenstates it was recently shown \cite{xxspin1} that the U(1) symmetric XX central spin model :
\bea
\hat{H} = B^z_0\hat{S}^z_0 +  \sum^{N-1}_{k = 1} 2 g_k \left(\hat{S}^x_0\hat{S}^x_k + \hat{S}^y_0\hat{S}^y_k\right)
\label{XXU1}
\eea
is actually integrable when the central spin is a spin-$1$. The spin$-1$ model also supports, for arbitrary coupling strengths, dark states such that the central spin is in either its lowest or highest weight state of the spin$-1$ representation $\left| S_0 = 1,m_z = \pm 1\right>$ having maximal or minimal eigenvalues $\pm1$ of the $\hat{S}^z_0$ operator.

Since the spin$-1$ model has not currently been shown to directly belong to a known class of R-G models, the question of whether it stays integrable or not in the presence of an in-plane magnetic field is worth investigating. Not only is that question relevant by itself, it can also circumscribe the candidates in the search for an explicit connection to R-G models. Indeed, it is known that most XXZ R-G models are exclusively  integrable in the presence of a z-oriented magnetic field, and only those which correspond to limits of the non-skew symmetric elliptic model support an in-plane component \cite{skrypnyk}. 
 
In this work, we first construct, in the next section, the conserved charges of the central spin$-1$ model in an arbitrarily oriented magnetic field, therefore proving its integrability. In the following section, polynomial relations between the conserved charges are built. In section \ref{darksec}, we then demonstrate numerically that, in the spin$-1$ case, dark states reemerge at strong coupling just as they do for central spin$-\frac{1}{2}$.
  
 \section{Conserved charges}
 
 In order to construct conserved charges which commute with one another and with the hamiltonian ($\ref{XXwithfield}$), we first choose to write it in terms of spin raising/lowering operators as:
\bea
\hat{H} &=& \vec{B} \cdot \vec{\hat{S}}_0 +  \sum^{N}_{k = 1} g_k \left(\hat{S}^+_0\hat{S}^-_k + \hat{S}^-_0\hat{S}^+_k\right)
\nonumber\\
&=& \vec{B} \cdot \vec{\hat{S}}_0 +  \left(\hat{S}^+_0\hat{G}^- + \hat{S}^-_0\hat{G}^+\right),
\label{XXwithfieldpm}
\eea
\noindent using the pair of hermitian conjugate operators:
\bea
\hat{G}^-  \equiv \sum^{N}_{k = 1} g_k \hat{S}^-_k
\ \ \ \ \ 
\hat{G}^+ \equiv \sum^{N}_{k = 1} g_k \hat{S}^+_k.
\eea
Using the explicit $3\times 3$ matrix representation, using the central spin's canonical basis of the three eigenstates of the $\hat{S}^z_0$ operator: $\left|S=1,m_z=1\right>,\left|S=1,m_z=0\right>\left|S=1,m_z=-1\right>$ which was also used in \cite{xxspin1}, one can write it conveniently as:
\bea
\hat{H} = 
 \left(
\begin{array}{ccc}
B_z &      \sqrt{2}\hat{\mathcal{G}}^-& 0  \\
   \sqrt{2} \hat{\mathcal{G}}^+ &  0  &      \sqrt{2} \hat{\mathcal{G}}^-\\
   0 &    \sqrt{2}\hat{\mathcal{G}}^+ &   - B_z
\end{array}
\right).
\nonumber\\
\label{XXwithfieldmatrix}
\eea
\noindent where we defined:
\bea
\hat{\mathcal{G}}^- & \equiv& \hat{G}^- + \frac{B_x - i B_y}{2}
 \nonumber\\ 
\hat{\mathcal{G}}^+ & \equiv& \hat{G}^+ + \frac{B_x + i B_y}{2}.
\eea
Using the same $3\times 3$ representation, one can build the following $N$ operators $\hat{R}_j  \ \ \ \forall j =1,2 \dots N$: 
\bea
\hat{R}_j = 
   \left(
\begin{array}{ccc}
B_z \left( \hat{\mathcal{Q}}_j - \hat{S}_j^z\right)&   \sqrt{2}\hat{\mathcal{G}}^-\hat{\mathcal{Q}}_j & 0  \\
 \sqrt{2}\hat{\mathcal{Q}}_j  \hat{\mathcal{G}}^+ & B_z \hat{S}_j^z  &   \sqrt{2}\hat{\mathcal{Q}}_j \hat{\mathcal{G}}^- \\
   0 &  \sqrt{2}\hat{\mathcal{G}}^+\hat{\mathcal{Q}}_j &   -  B_z \left( \hat{\mathcal{Q}}_j + \hat{S}_j^z\right)
\end{array}
\right).
\nonumber\\
\label{Rjmatrix}
\eea
\noindent where:
\bea
 \hat{\mathcal{Q}}_j  & \equiv&  \hat{Q}_j + \frac{B_x \hat{S}^x_j + B_y \hat{S}^y_j}{g_j}
\eea
\noindent and 
\bea
&& \hat{Q}_j = \frac{\hat{S}^+_j\hat{S}^-_j +\hat{S}^-_j\hat{S}^+_j}{2}  \nonumber\\ &&+ \sum_{k=1 (\ne j)}^N \left[\frac{g_j g_k}{g_j^2 - g_k^2}\left(\hat{S}^+_j\hat{S}^-_k +\hat{S}^-_j\hat{S}^+_k\right)+\frac{2 g_k}{g_j^2 - g_k^2}\hat{S}^z_j\hat{S}^z_k \right].\nonumber\\
\eea
The operators (\ref{Rjmatrix}) precisely correspond to the form of the conserved charges found without an in-plane field \cite{xxspin1}, but modified by the substitutions $ \hat{Q}_j  \to  \hat{\mathcal{Q}}_j$ and  $\hat{G}^\pm \to \hat{\mathcal{G}}^\pm$. In order to compute the required commutators, one can first show, using the fact\cite{lukya} that   $[\hat{Q}_i,\hat{Q}_j]=0$, that $[\hat{\mathcal{Q}}_i,\hat{\mathcal{Q}}_j]=0$ commute with one another. One can then prove that the relation $\left[\hat{\mathcal{G}}^\pm,\hat{Q}_j\right] = \pm\left\{ \hat{S}^z_j,\hat{\mathcal{G}}^\pm \right\} \pm \left(B_x\pm i B_y\right)\ \hat{S}^z_j$ holds. It is then relatively straightforward, but tedious, to show that the conserved charges all commute with  the Hamiltoninan $\left[\hat{R}_j,\hat{H}\right] = 0$. Showing that they also commute with one another $\left[\hat{R}_j,\hat{R}_k\right] = 0$ is also easily achieved through direct calculation. 

These operators therefore form a set of conserved charges whose existence are sufficient \cite{mossel} to conclude that the XX spin-1 central spin model remains integrable in the presence of an arbitrarily oriented magnetic field. This can certainly have an important impact on the way the model could relate to known classes of R-G models since, as mentioned in the introduction, only a limited class of models remain integrable in the presence of transverse magnetic field components.

 \section{Polynomial relations}
 
The set of conserved charges  (\ref{Rjmatrix}) proposed in this work and the hamiltonian (\ref{XXwithfieldmatrix}), also obey a set of $N+1$ polynomial equations relating the various operators. The construction is similar to the quadratic equations one finds for the typical R-G systems built out of only spin-$\frac{1}{2}$ realisations \cite{dimoquad, skrypnykquad}. However for a central spin$-1$, one finds the cubic relation: 
\bea
\hat{H}^3 = \sum_{j=1}^N 4 g_j^2 \hat{R}_j + \left|B\right|^2 \hat{H}.
\label{cubic}
\eea
\noindent Once again it can be proven, by direct calculation using the $3 \times 3$ matrix representation specific to the central spin$-1$ problem. A useful intermediate result for this proof is to first show that $\left(\hat{\mathcal{G}}^+\hat{\mathcal{G}}^-+\hat{\mathcal{G}}^-\hat{\mathcal{G}}^+ \right) - \frac{(B_0^x)^2+(B_0^y)^2}{2} \hat{\mathbb{1}}  = 2 \displaystyle \sum_{i=1}^N g_i^2 \hat{\mathcal{Q}}_i$.  As was the case for the conserved charges, the proof of this cubic relation (\ref{cubic}) relies exclusively on the $SU(2)$ commutation relations $\left[\hat{S}^\alpha_j,\hat{S}^\beta_k\right] = i \delta_{jk}\epsilon_{\alpha,\beta,\gamma} \hat{S}^\gamma_j $ for the bath spins. It, therefore, is valid for arbitrary representations of the various environmental spins (i.e. be they, individually, spin$-\frac{1}{2},\ -1, \ -\frac{3}{2}, \dots$). 

In order to form a complete set of $N+1$ polynomial relations, we now choose, for simplicity, the bath spins to all be spins-$\frac{1}{2}$. In doing so the $N$ remaining polynomial relations are simply quadratic and are explicitly given by:

\bea
 && \hat{R}_j^2 = \frac{B_z^2}{4} \hat{\mathbb{1}} +   \hat{H} \hat{R}_j 
  +  \sum_{k \ne j}^N \left(\frac{g_k^2}{g_j^2 - g_k^2}\right)  \hat{H} \hat{R}_k \nonumber\\ && \ \ \ 
- \left(\frac{1}{4} - \frac{B_x^2+B_y^2}{4g_j^2} - \sum_{k \ne j}^N \left[\frac{3}{4}\frac{g_k^4}{\left(g_j^2 - g_k^2\right)^2}\right] \right) \hat{H}^2,
\nonumber\\
\eea

\noindent once again a result which is provable by direct calculation of the square and product of the involved operators. To do so one can first compute the square of the $\hat{\mathcal{Q}}_i$ operator and use the specificities of spin$-\frac{1}{2}$ Pauli matrices (such as $\left(S^\pm\right)^2 = 0$) to prove the equality. 

Polynomial relations between conserved operators have formed the basis of the so-called eigenvalue-based approach to R-G models. Here, one can expect, as it is the case with the usual R-G models, to be able to write $R_j^{2S_j+1}$ in terms of lower/equal powers of the other conserved charges: a cubic relation for spin-$1$, a quartic one for spin-$\frac{3}{2}$, etc. In \cite{linkstalk} one can find an explicit example for the XXX spin-1 model, while for higher-spin XXX models one can infer such relations using known results for polynomial constructions linking eigenvalues and their derivatives \cite{araby}. However, only the specific case of a bath of spins-$\frac{1}{2}$, shown in the previous equation, will be explicitly constructed in this work.

Since all these operators are diagonal in the basis formed by their common eigenstates:
\bea
\hat{H}\left| \psi_n\right> = E^n_0 \left| \psi_n\right> \ \ \ \ \ \hat{R}_j\left| \psi_n\right> = E^n_j \left| \psi_n\right>,
\eea

\noindent the same polynomial relations are also valid for the set of eigenvalues associated to any eigenstate $\left| \psi_n\right>$:
\bea
&& \left(E^n_0\right)^3 = \sum_{j=1}^N 4 g_j^2 E^n_j + \left|B\right|^2 E^n_0
\nonumber\\
 &&\left(E^n_j\right)^2 = \frac{B_z^2}{4} +   E^n_0E^n_j
  +  \sum_{k \ne j}^N \left(\frac{g_k^2}{g_j^2 - g_k^2}\right)  E^n_0 E^n_k \nonumber\\ && \ \ \ 
- \left(\frac{1}{4} - \frac{B_x^2+B_y^2}{4g_j^2} - \sum_{k \ne j}^N \left[\frac{3}{4}\frac{g_k^4}{\left(g_j^2 - g_k^2\right)^2}\right] \right) \left(E^n_0\right)^2. 
\nonumber\\
\label{eigenvaluesystem}
\eea

One can therefore circumvent the Bethe Ansatz and avoid looking for rapidities (Bethe roots)  which are solution to Bethe equations. One can instead, find the solutions to the previous polynomial equations directly giving the eigenvalues $(E^n_0,E^n_1,E^n_2 \dots E^n_N)$ associated to any given eigenstate $\left|\psi_n\right>$. In many R-G models, it then becomes possible to build determinants expressions which can give access to scalar products and matrix elements of various local operators directly in terms of these eigenvalues \cite{faridet,claeysdet,claeysdet2,johnsondet}. However, in the specific case at hand we will simply use the Hellmann-Feynman theorem in order to study, in the next section, the  physical properties of the central spin, in various eigenstates, as a function of the intensity of the coupling.

\section{Dark states}
\label{darksec}

As mentioned in the introduction, the existence of dark states is a defining feature of both the spin-$\frac{1}{2}$ model \cite{XXdark} and the spin-$1$ model\cite{xxspin1} in the $U(1)$ symmetric case, i.e.: with a z-oriented  magnetic field. In both cases it was shown that there exists a class of eigenstates, for which the central spin is in a pure state completely unentangled with the bath. In both cases, they occur for a central spin in either of the highest or lowest weight state of the representation, namely $\left| S_0 = \frac{1}{2},m_z = \pm \frac{1}{2}\right>$ or $\left| S_0 = 1,m_z = \pm1\right>$. These dark states exist for arbitrary couplings and stay unentangled when the overall coupling strength is raised or lowered. One can easily get an intuitive physical picture justifying the existence of such states: 
\bea
\left|\psi_\mathrm{dark}\right> =  \left| S_0, m_z = \pm S_0\right>\otimes\left|\psi_\mathrm{bath}\right>.
\eea

\noindent In the highest and lowest weight states, the central spin is perfectly aligned with the $z$ axis and therefore with the magnetic field making it an eigenstate of the magnetic part of the hamiltonian $B_0^z \hat{S}^z_0$. Considering that the coupling term only involves the XY components of the central spin, it then seems perfectly reasonable that a z-axis aligned central spin can stay completely unentangled with the bath since it points in a direction along which it does not couple to the bath at all.

In light of this simple picture, it is somehow surprising that, in the central spin-$\frac{1}{2}$ case it was shown that  dark states can also emerge\cite{dimodark}. However, they do so through the reorganisation of the bath spins, and appear exclusively when the overall coupling strength becomes strong enough \cite{dimodark}. The strong coupling emergence of dark states was understood as the result of the Overhauser effective magnetic field due to the bath: $\sum_{j=1}^N g_j \left< \vec{S}_j\right>$, exactly cancelling  the external in-plane magnetic field components and effectively bringing the problem back to the U(1) symmetric case. Since the resulting effective field is proportional to the overall coupling strength, this cancellation does require a strong enough coupling to occur.

For spin-$\frac{1}{2}$ the question of whether the central spin is in a pure state or not  was easily addressed. Since the condition $ \left<S^z_0\right>^2+\left<S^y_0\right>^2+\left<S^x_0\right>^2 = \frac{1}{4}$ is equivalent to having the reduced density matrix of the central spin $\rho_0 = \frac{1}{2} + \vec{n}\cdot\vec{\sigma}$  defined by a vector $\vec{n}$ with norm $1$ and therefore on the surface of the Bloch sphere, it is completely sufficient to signal a pure state. Any mixed state would be represented by a vector of smaller norm, lying inside the Bloch sphere, and characterised by $ \left<S^z_0\right>^2+\left<S^y_0\right>^2+\left<S^x_0\right>^2 < \frac{1}{4}$ . The generalisation to spin-$1$ is much more complicated since a general density matrix is then defined by the Gell-Mann matrices and an 8-dimensional vector for which two conditions have to be met to define a pure state (one on the norm and one on the so-called star product) \cite{blochspin1}. In this work, a complete verification of whether a given eigenstate is on, or close to, the generalized Bloch sphere is therefore a much more complicated task. It would, at least, require an explicit construction of the eigenstates using a Bethe Ansatz which, for the time being, has not yet been built for the specific problem of the XX model in a generic magnetic field. Nonetheless, considering the way spin-$\frac{1}{2}$ dark states emerged at strong coupling with an in-plane field, one can here simply look at the average of the $S^z_0$ component. If its average is $\pm 1$, the state of the central spin will then assuredly be the pure state $\left|S=1, m=\pm1\right>$ insuring that it has no entanglement with the bath.  Indeed, for any state where the reduced density matrix of the central spin is in a mixed state, the expectation value $\left<S^z_0\right>$ will have a norm which is lower that 1. Evidently, it would still be possible for the central spin to be in a pure state without having $\left<S^z_0\right> = \pm 1$ but metting this condition is sufficient to show that the central spin is in a pure state and that therefore the system is in an unentangled dark state.

Computing $\left<S^z_0\right>$ numerically is very simple for any eigenstate of the system since one will use the system of polynomial equations (\ref{eigenvaluesystem}) to explicitely find one (or many) solution giving the ensemble of $E^n_j$ eigenvalues which define a particular eigenstate. 
The derivatives of the polynomial system, with respect to any parameter, provides a set of linear equations, for the derivatives of the eigenvalues, whose coefficients are known when the ensemble of eigenvalues is known. Through the Hellmann-Feynman theorem, these derivatives can be directly related to quantum expectation values of operators in that particular eigenstate. In the case at hand:
\bea
 \left<\psi_n\right|\hat{S}^z_0\left|\psi_n\right>=   \left<\psi_n\right|\frac{\partial \hat{H} }{\partial B_z}\left|\psi_n\right> = \frac{\partial E^n_0 }{\partial B_z}.
\eea

\noindent which, using the compact notation $E'^n_j \equiv \frac{\partial E^n_j}{\partial B_z }$, can be computed by solving the linear system found by deriving (\ref{eigenvaluesystem}) with respect to $B_z$:

\bea
&& 
3 \left(E^n_0\right)^2  E'^n_0= \sum_{j=1}^N 4 g_j^2 E'^n_j + \left|B\right|^2 E'^n_0 + 2 B_z E^n_0
\nonumber\\
 &&2 \left(E^n_j\right) E'^n_j =   \frac{B_z}{2}+ E'^n_0E^n_j + E^n_0 E'^n_j\nonumber\\ && \ 
  +  \sum_{k \ne j}^N \left(\frac{g_k^2}{g_j^2 - g_k^2}\right)  \left( E'^n_0 E^n_k +E^n_0 E'^n_k \right)\nonumber\\ && \ 
- 2 E^n_0 E'^n_0 \left(\frac{1}{4} - \frac{B_x^2+B_y^2}{4g_j^2} - \sum_{k \ne j}^N \left[\frac{3}{4}\frac{g_k^4}{\left(g_j^2 - g_k^2\right)^2}\right] \right) . 
\nonumber\\
\label{eprimeystem}
\eea

In a way which has become standard practice \cite{araby,solver,dimoclaeys,johnson}, we will find individual solutions, one by one, by deforming one of the known solutions at zero-coupling. Defining $g_k = g \epsilon_k$,  $g$ will be varied continuously from zero to the large coupling limit.  In the non-interacting limit $g\to 0$ the first equation of (\ref{eigenvaluesystem}) reduces to:
\bea
&& \left(E^n_0\right)^3 =  \left|B\right|^2 E^n_0
\eea
\noindent and therefore to three, massively degenerate, eigenvalues:
\bea
E^n_0(g=0) &\in& \{-\left|B\right|,0,\left|B\right|\} .
\eea
In the two cases, where $E^n_0 \ne 0$, the other equations of (\ref{eigenvaluesystem}) reduce to:
\bea
\nonumber\\
 &&\left(g E^n_j\right)^2 = \frac{B_x^2+B_y^2}{4 \epsilon_j}  \left(E^n_0\right)^2. 
\nonumber\\
\eea
\noindent leading to the set of $g=0$ solutions given by 
\bea
E^n_0(g=0) &=& \pm \left|B\right| \ ; \ \tilde{E}^n_j(g=0) = \pm \frac{ \sqrt{B_x^2+B_y^2}}{2 \epsilon_j  } \left|B\right|  
\nonumber\\
\eea
\noindent  with $\tilde{E}^n_j \equiv g E^n_j$. 
The remaining solutions are found for $E^n_0(g=0) = 0$ and, since one can easily show that $E^n_0(g \to 0) \propto g^2$, the last equations simply become $E^n_j(g=0)  = \frac{B_z^2}{4}$. We therefore have:
\bea
E^n_0(g=0) &=& 0 \ ;\ E^n_j(g=0) = \pm \frac{ B_z}{2},
\eea
\noindent which completes the list of possible $g=0$ solutions.  Starting from any of these solutions, one can then simply use the Newton-Raphson method to solve the system at a small finite $\delta g$ using the $g=0$ as the starting point of the iterative process. Repeating the process by using the solution at the current $g$, or a better approximation built using the Taylor series \cite{araby,solver}, as the starting point for solving at $g+\delta g$ ultimately allows one to deform any $g=0$ solution into one unique solution over a chosen range of coupling intensities.

\begin{figure}[htbp]
\begin{center}
\includegraphics[width=9.5cm]{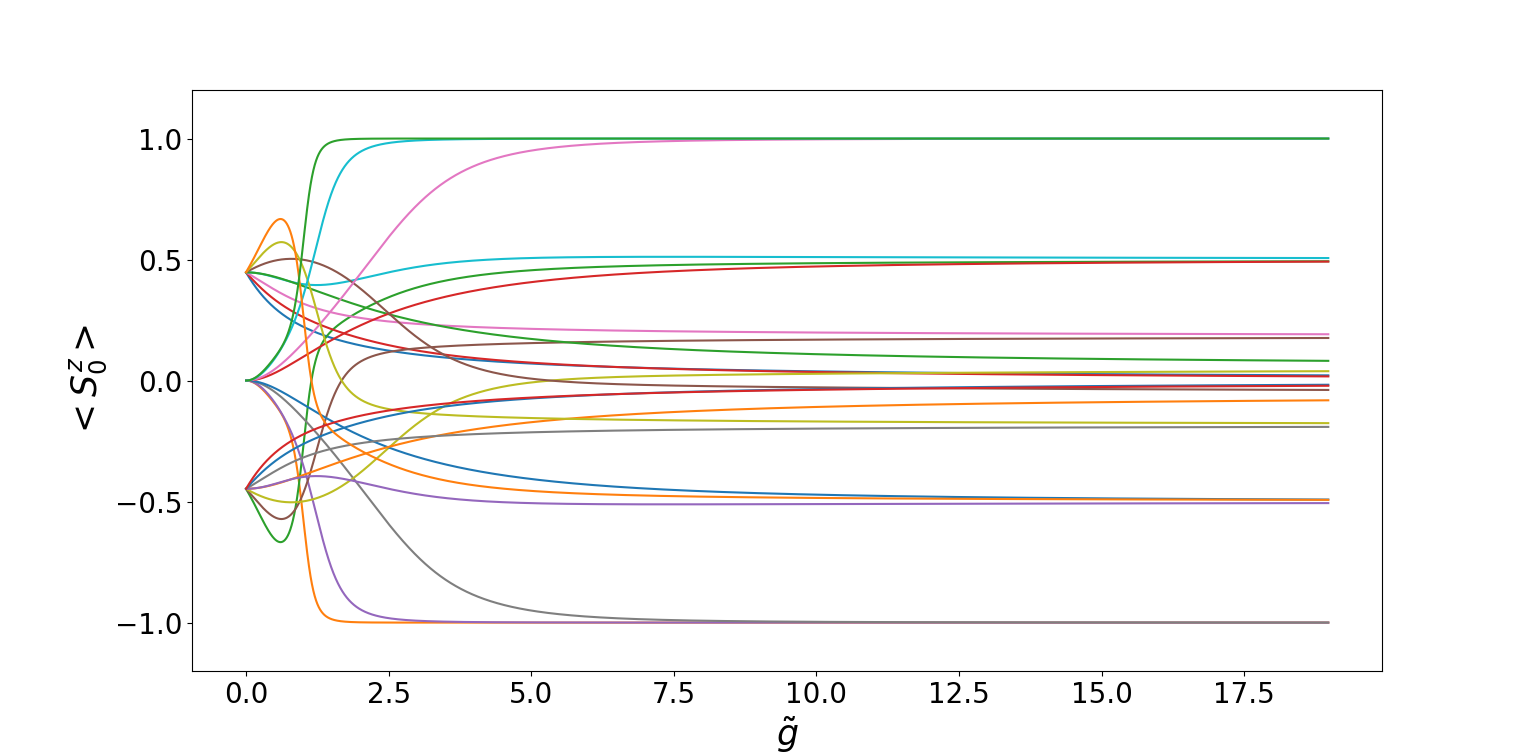}
\caption{Central spin's z-projection expectation value as a function of the renormalised coupling $\displaystyle \tilde{g} = \frac{\sum_{k=1}^N g \epsilon_k}{\left|B\right|}$ for the full Hilbert space with $N=3$ bath spins. The parameters are chosen as $\epsilon_k = k$. The orientation of the field is chosen so that $B_x = 2 B_z$.}
\label{allstates}
\end{center}
\end{figure}

In Fig. \ref{allstates}, we plot the expectation of $S^z_0$ for the complete set of eigenstates of a small $N=3$ system, as a function of the overall coupling strength $g$, after renormalising it to $\displaystyle \tilde{g} = \frac{\sum_{k=1}^N g \epsilon_k}{\left|B\right|}$.

It is clear that a fraction of the states ( 6 out the 24 states in this specific case) become, at strong enough coupling, dark states. Indeed, the two set of 3 states for which the central spins is either pointing "fully up" or "fully down" ($\left<S^z_0\right> = \pm 1$) along the z axis are unavoidably such that the central spin is the pure state: $\left|S=1,m_z=\pm1\right>$. It is essential to notice that the renormalised coupling at which the entanglement disappear is such that the total coupling and magnetic field are of the same order of magnitude. It therefore occurs far from the point at which the coupling would be sufficiently strong to completely neglect the external magnetic field. This indicates that, as was the case for a central spin$-\frac{1}{2}$, it is the effective magnetic field created by the arrangement of the bath spins which allows dark states to reemerge at such coupling strengths where $\frac{1}{|B|}$ is completely non-perturbative.

\begin{figure}[htbp]
\begin{center}
\includegraphics[width=9.5cm]{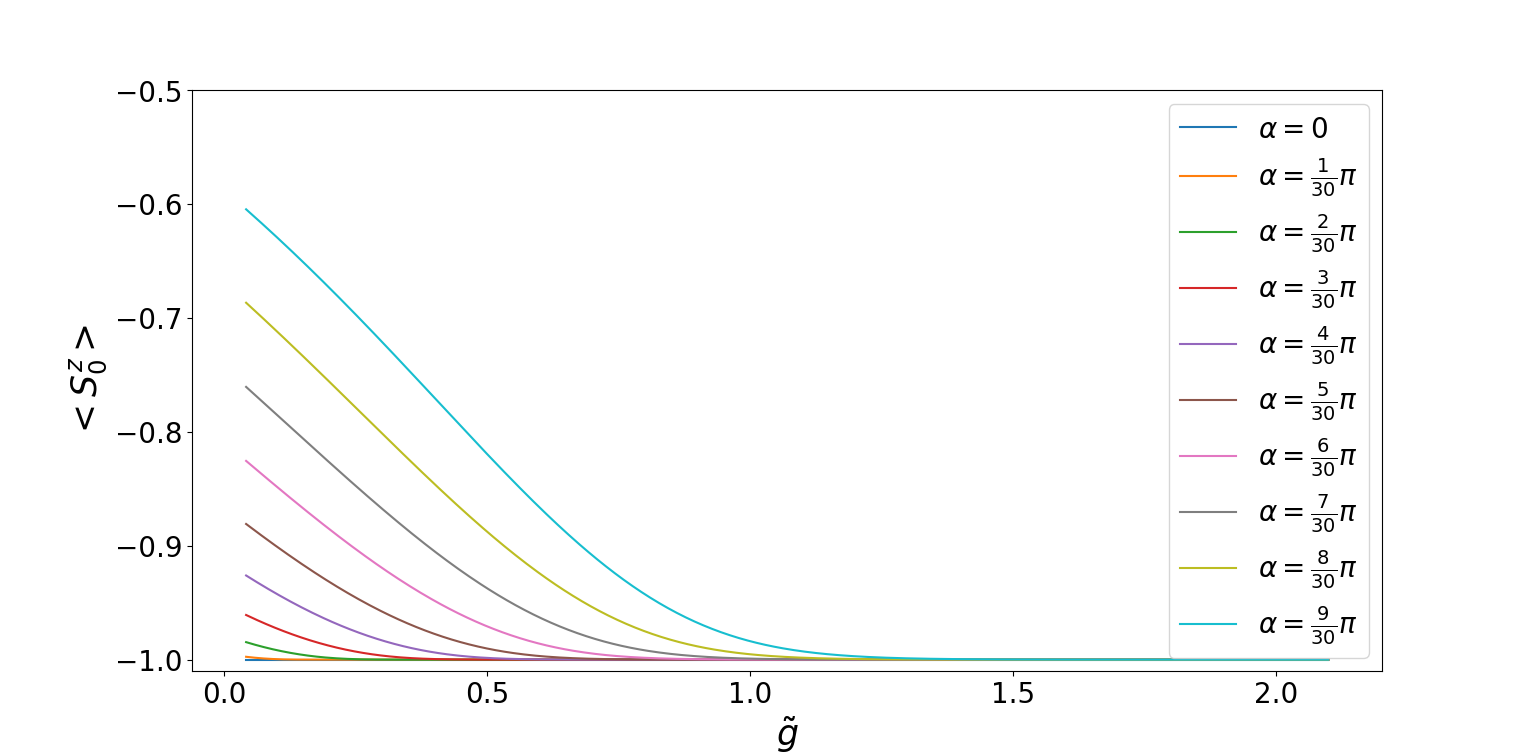}
\caption{Central spin's z-projection expectation value as a function of the renormalised coupling $\tilde{g} = \frac{\sum_{k=1}^N g \epsilon_k}{\left|B\right|}$ for a specific dark state with different angles $\cos{\alpha} = \frac{B_z}{\left|B\right|}$. The results are for $N=20$ bath spins and the parameters are chosen as $\epsilon_k = k$}
\label{angles}
\end{center}
\end{figure}

Going to a larger system of $N=20$ bath spins, we choose to look at one specific eigenstate which results from the deformation of the $g=0$ configuration with $E_0 = -\left|B\right|$,  $E_i = - \frac{ \sqrt{B_x^2+B_y^2}}{2 \epsilon_j  } \left|B\right|$ for  $i = 1,3,6,8$ and  $+ \frac{ \sqrt{B_x^2+B_y^2}}{2 \epsilon_j  } \left|B\right| $ for the remaining  values of $i$. Changing the orientation of the field through an azimuthal tilt, i.e. $B_z = |B| \cos(\alpha)$  (XY-plane component given by $|B| \sin(\alpha)$) leads to figure \ref{angles} when plotting the expectation value of the central spin as a function of the renormalised coupling.

These plots first demonstrates that the presence of dark states in the strong coupling limit was not due to the tiny system size presented before since the dark state feature is clearly seen emerging at strong coupling again in this larger system. Moreover, these results make it clear that it is not the magnitude of the magnetic field, but exclusively the magnitude of its in-plane component which controls the coupling strength necessary for the dark state to reemerge. The result is completely consistent with the mechanism which was at play for central spin$-\frac{1}{2}$. Despite the fact that we are not in a position to explicitly compute the expectation values of the various bath spins to confirm it explicitly, the observed behaviour of the central spin strongly support the hypothesis that similar physics is at play. Indeed the whole phenomenology observed here is precisely the same as what was seen in the spin$-\frac{1}{2}$ case:
the system reproduces the U(1)-symmetric dark states, does so exclusively when the coupling gets strong enough and the required coupling strength is controlled by the value of the in-plane magnetic field component which has to be cancelled to restore an effective U(1)-symmetry.

\section{Conclusion}

In this work, we have shown that the integrability of the XX central spin-1 model, is maintained in the presence of an arbitrarily oriented magnetic field which breaks the rotational U(1)-symmetry around the z axis. 

The set of commuting conserved charges and the hamiltonian have been shown to obey polynomial equations, computed here when the bath spins are all spin$-\frac{1}{2}$, which can be used in order to numerically access physical properties of its eigenstates. Doing so, we have explicitly shown that dark states, for which the central spin is completely unentangled with the bath do exist in this system. However, they only emerge, as was the case when the central spin is spin$-\frac{1}{2}$, when the coupling is strong enough.  

It remains to be seen explicitly if the integrability of the central spin model is true irrespective of the central spin's realisation. This work seems to suggest that if it exists, the connection to a R-G-like construction could be made the models which are not U(1)-symmetric. Being able to make such a connection could allow proofs purely based on algebraic considerations which, being independent of the realisation, would generalise to arbitrary central-spins.

\newpage

\

\end{document}